# Improving the convergence of coupled quasinormal-mode theory for optical resonators

Tong Wu[1,2], Philippe Lalanne[3†], Haitao Liu[1,2*]

[1]Institute of Modern Optics, College of Electronic Information and Optical Engineering, Nankai University, Tianjin 300350, China

[2]National Key Laboratory of Semiconductor Laser, Tianjin Key Laboratory of Micro-scale Optical Information Science and Technology, Tianjin 300350, China

[3]LP2N, CNRS, IOGS, Université Bordeaux, Talence 33400, France

[†]philippe.lalanne@institutoptique.fr

[*]liuht@nankai.edu.cn

## Abstract

Quasinormal modes (QNMs) provide a rigorous framework for describing resonances in open optical systems, including radiative and material losses. Recent coupled-QNM theories use the modes of individual resonators to predict the resonances and optical responses of coupled systems, without relying on phenomenological coupling parameters. However, their accuracy can deteriorate when the resonators confine light weakly and radiate strongly, often requiring many modes to obtain reliable predictions. Here, we introduce a reformulation of coupled-QNM theory that substantially improves its accuracy when only a small number of modes of the individual resonators are retained. The approach builds on the same underlying QNM framework but treats the fields mediating the interaction between resonators differently, leading to a new expression for the coupling coefficients. Numerical comparisons show that this extended-domain formulation is particularly advantageous for resonators with strong radiative losses, enabling more efficient modeling of a broader range of coupled photonic systems.

## 1. Introduction

The interaction between optical resonators provides a powerful way to control light at the nanoscale [1-5]. When two or more resonators are brought together, their individual modes combine to form hybrid states, or *supermodes*, whose properties can differ markedly from those of the isolated resonators. Their resonance frequencies, lifetimes, near-field distributions, and radiation patterns can all be modified. This ability underlies a wide range of phenomena and applications, including chiral resonant states [6,7], level attraction [8,9], subwavelength cavity modes with exceptionally high quality factors [10,11], and parity-time (PT)- and anti-PT-symmetric systems [12,13].

Predicting these interactions quantitatively remains challenging, particularly when radiation and material absorption play an important role. Conventional coupled-mode theory (CMT) provides an intuitive and versatile description of interacting resonators [14,15]. However, its coupling coefficients are often introduced as phenomenological parameters fitted to numerical or experimental results. This limits the model's predictive power and can become problematic in strongly coupled or strongly dissipative systems, where assumptions such as weak coupling, high quality factors, and modal orthogonality may no longer hold [16].

A more rigorous approach is to derive the interaction directly from Maxwell's equations, using the electromagnetic properties of the individual resonators as the starting point. Quasinormal-mode (QNM) theory offers a natural foundation for this strategy. QNMs describe the resonances of open electromagnetic systems, accounting for their leakage of energy into the surrounding environment, and the theory has been extended to accommodate material loss and dispersion [17]. Recent studies have therefore developed coupled-QNM (cQNM) formulations that use the modes of individual resonators to predict the resonances and optical responses of the combined system [18-26]. By deriving the coupling coefficients from Maxwell's equations, these methods can predict the hybrid modes without fitting phenomenological parameters.

Despite this progress, obtaining accurate predictions with a manageable number of QNMs remains a central challenge. One difficulty stems from the distinctive behavior of QNMs in open systems: their radiative fields grow exponentially with distance from the resonator [27]. This spatial divergence complicates the representation of fields outside the resonator and can lead to counterintuitive, but now established, behavior, such as the exponential growth of coupling coefficients as the separation between resonators increases [24,27]. The growth is actually what distinguishes dissipative (non-Hermitian) coupling between open-resonators from the coupling between Hermitian modes in closed systems.

Material dispersion introduces an additional complication. Because the material response depends on frequency, the coupled-mode equations can become nonlinear in the unknown eigenfrequency, making the calculation of supermodes more involved [26]. Existing cQNM formulations address these challenges in different ways, using various treatments of QNM exponential growth, normalization, material dispersion, and field reconstruction. Their respective approximations and ranges of validity can differ, and rigorous demonstrations of their accuracy remain scarcely considered in the literature [23,26].

A recent study [23] proposed a systematic comparison of existing approaches and their respective capabilities. It also developed a cQNM framework that addresses several of these issues. The formulation derives coupling coefficients directly from Maxwell's equations, accounts for material dispersion through auxiliary fields, and uses perfectly matched layer (PML) regularization to handle the spatial divergence of QNMs. It also yields an evolution equation in which the modal excitation coefficients and their time derivatives both contribute to the coupling dynamics. Numerical tests demonstrated its applicability to dispersive materials and strongly interacting resonators, including systems with weakly confined fields. Beyond providing an efficient means of predicting coupled resonances, the framework offers a solid interpretation of mode couplings level-attraction and large imaginary mode volumes.

However, the efficiency of this formulation depends on the optical properties of the individual resonators. High accuracy can often be obtained with only a few QNMs for systems with strong field confinement, such as subwavelength plasmonic antennas and photonic microcavities. For resonators with weaker confinement and stronger radiation losses [28-30], convergence can be substantially slower, so that many more modes must be included to achieve comparable accuracy [23]. Reducing this modal requirement would make coupled-QNM theory more efficient and extend its practical usefulness to a broader range of photonic systems.

Here, we address this limitation by reformulating the derivation of the cQNM equations. The new approach retains the general structure of the previous evolution equation, including coupling terms involving time derivatives of the modal excitation coefficients, but leads to different coupling coefficients. The key difference lies in how the fields mediating the interaction between resonators are represented. In the earlier scattered-domain formulation (SCAdom) [23], the field scattered by each individual resonator is expanded separately in the resonator's QNM basis, and the resulting coupling integrals are evaluated over the individual resonator volumes. In the present extended-domain formulation (EXTdom), the fields scattered by every individual resonator is no longer central. The field scattered by all resonators is instead represented using the combined set of QNMs from all individual resonators. The corresponding overlap integrals extend across the entire regularized computational domain, including the PML regions.

This change in formulation leads to a substantial improvement in accuracy when only a small number of QNMs is retained. The advantage is particularly pronounced for resonators with strong radiative losses, for which the SCAdom formulation may require many modes to achieve reliable predictions. The precise origin of this improved convergence with a limited number of modes is not yet fully understood. One possible explanation is that the EXTdom integrals preserve information about the interaction that is not captured when the integrals are restricted to the individual resonator volumes. The two formulations also arise from different mathematical approaches: the EXTdom formulation belongs to the broader mathematical tradition of Rayleigh–Ritz method [31], whereas the SCAdom formulation is based on an ad hoc multiple-scattering theory. These differences may contribute to their distinct convergence behavior.

The following sections first derive the EXTdom formulation for systems with dispersive or non-dispersive materials, with frequency-independent coupling coefficients. We then compare its accuracy with that of the early SCAdom formulation for both eigenfrequency prediction and optical-response calculations, examining how the results depend on the number of QNMs retained. Finally, we address a practical limitation of the EXTdom approach and show how this difficulty can be overcome by replacing integrals over the entire regularized numerical space with integrals evaluated on or within the resonator boundaries.

## 2. Extended-domain cQNM theory

As shown in Figure 1, consider two coupled optical resonators, *A* and *B*, with a straightforward extension to multiple resonators. We adopt the conventional scattered-field formulation, in which the permittivity is written as $\varepsilon(\mathbf{r},\omega) = \varepsilon_b(\mathbf{r},\omega) + \Delta\varepsilon^A(\mathbf{r},\omega) + \Delta\varepsilon^B(\mathbf{r},\omega)$, where $\Delta\varepsilon^X$ with $X \in \{A,B\}$ denotes the permittivity change introduced by the resonator, and $\varepsilon_b$ is the permittivity of the background medium. The coupled system is illuminated by a monochromatic driving field $\mathbf{E}_D$, defined as the field in the absence of both resonators (i.e., in the background medium).

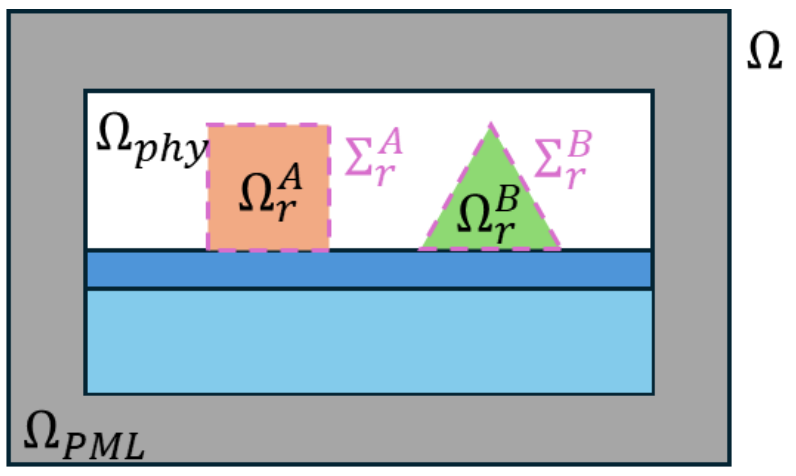

**Figure 1 Illustration of the coupled system under study.** The coupled system consists of two nearby resonators, labeled *A* or *B*, with permittivity changes $\Delta\varepsilon^A$ and $\Delta\varepsilon^B$, respectively, relative to the background permittivity $\varepsilon_b$ within the domains $\Omega_r^A$ and $\Omega_r^B$. After a complex coordinate transformation is applied for regularization, the mapped computational space is divided into an exterior PML domain $\Omega_{PML}$, shown in gray, and an inner physical domain $\Omega_{phy}$ containing the two resonators. The full computational domain is $\Omega = \Omega_{PML} \cup \Omega_{phy}$. $\Omega_c$ in eqs. 4 and 5 is defined as $\Omega_c = \Omega - \Omega_r^A - \Omega_r^B$ and represents the domain out of the resonators. The boundaries of the resonators are denoted by $\Sigma_r^A$ and $\Sigma_r^B$.

To ensure analytical continuation of the permittivity into the complex-frequency plane, we assume that every dispersive material follows a standard single-pole Lorentz model, $\varepsilon(\omega) = \varepsilon_\infty - \varepsilon_\infty \frac{\omega_p^2}{\omega^2 + i\omega\gamma - \omega_0^2}$, where $\omega_p$, $\gamma$, and $\omega_0$ are the plasma frequency, damping coefficient, and material resonance frequency, respectively. The generalization of the following theory to an *N*-pole Lorentz permittivity is straightforward.

The field $\boldsymbol{\psi}$ scattered by the coupled system excited by a source is governed by Maxwell's equations which can be recast in the form [32]

$$\mathbf{H}\boldsymbol{\psi}(\mathbf{r},\omega) = \omega\boldsymbol{\psi}(\mathbf{r},\omega) + \mathbf{S}(\mathbf{r},\omega), \quad (1)$$

where $\mathbf{H}$ is a Hamiltonian operator independent of $\omega$ determined by the intrinsic properties of the coupled system, namely, the spatial distribution of the permittivity. The term $\mathbf{S}$ represents the source term associated with the external driving field $\mathbf{E}_D$. Their explicit expressions are given in Supporting Information, S1. $\boldsymbol{\psi} = \left[\varepsilon_\infty^{1/2}\mathbf{E}_s, i\mu_0^{1/2}\mathbf{H}_s, \omega_0\omega_p^{-1}\varepsilon_\infty^{-1/2}\mathbf{P}', -i\omega_p^{-1}\varepsilon_\infty^{-1/2}\mathbf{J}'\right]^T$ is an augmented field comprising the electric and magnetic fields together with auxiliary fields introduced to linearize the frequency dependence of Maxwell's equations. In particular, $\mathbf{E}_s$ and $\mathbf{H}_s$ are the scattered fields, while the auxiliary fields are defined as $\mathbf{P}' = [\varepsilon(\mathbf{r},\omega) - \varepsilon_\infty(\mathbf{r})]\mathbf{E}'$ and $\mathbf{J}' = -i\omega\mathbf{P}'$, where $\mathbf{E}' = \mathbf{E}_{tot} = \mathbf{E}_s + \mathbf{E}_D$ takes the total electric field for $\mathbf{r} \in \Omega_r^A \cup \Omega_r^B$ and $\mathbf{E}' = \mathbf{E}_s$ for $\mathbf{r} \in \Omega_c = \Omega - \Omega_r^A - \Omega_r^B$. The prefactors appearing in $\boldsymbol{\psi}$, such as $\varepsilon_\infty^{1/2}$ and $i\mu_0^{1/2}$, are introduced so that operator $\mathbf{H}$ is symmetric, making the expressions of QNM normalization and orthogonality more compact. In [18], a cQNM formulation was developed for nondispersive resonators. In the present work, we extend this theory to dispersive systems.

Hereafter, we employ a regularized-QNM framework based on a rotation of the complex-frequency plane using PMLs [17,33,34]. This transformation maps the exponentially divergent QNMs of open systems into square-integrable wavefunctions, enabling a well-defined modal expansion. Define augmented normalized modal fields of subsystems $A$ and $B$ as, $\tilde{\boldsymbol{\psi}}_m^A(\mathbf{r})$ and $\tilde{\boldsymbol{\psi}}_m^B(\mathbf{r})$, which consist of both dominant QNMs and PML modes. The former describe the resonant response of the subsystem, while the latter ensure the completeness of the modal expansion over the entire computational domain. The specific expressions of $\tilde{\boldsymbol{\psi}}_m^A$ and $\tilde{\boldsymbol{\psi}}_m^B$ and their normalization and orthogonality relationships are given in Supporting Information, S1.

For a system composed of two resonators, the scattered field $\boldsymbol{\psi}$ can be expressed as the sum of the scattered fields associated with the two individual resonators. Each scattered field satisfies the corresponding Maxwell's equation, with a source term determined by the excitation $\mathbf{S}$ and the field scattered by the other resonator (see Supporting Information, S2 for details). Since the QNMs and PML modes of each individual resonator form a complete basis, $\boldsymbol{\psi}$ can be expanded as:

$$\boldsymbol{\psi}(\mathbf{r},\omega) = \sum_m[\alpha_m^A(\omega)\tilde{\boldsymbol{\psi}}_m^A(\mathbf{r}) + \alpha_m^B(\omega)\,\tilde{\boldsymbol{\psi}}_m^B(\mathbf{r})], \tag{2}$$

with $\alpha_m^A$ and $\alpha_m^B$ the weight factors and $m = 1,2,\ldots$ Substituting eq 2 into eq 1 and projecting it onto the same basis using the orthogonality relation [32], one can obtain the following cQNM equation:

$$\hat{\mathbf{H}}|\alpha(\omega)\rangle = \omega\hat{\mathbf{D}}|\alpha(\omega)\rangle + |\mathbf{F}(\omega)\rangle \tag{3}$$

with $|\alpha(\omega)\rangle = [\alpha_1^A, \alpha_2^A, \ldots, \alpha_1^B, \alpha_2^B, \ldots]^T$ a vector formed by the weight factors, $\hat{\mathbf{H}} = \begin{bmatrix} \tilde{\boldsymbol{\omega}}^A & \boldsymbol{\kappa}^{AB} \\ \boldsymbol{\kappa}^{BA} & \tilde{\boldsymbol{\omega}}^B \end{bmatrix}$, $\hat{\mathbf{D}} = \begin{bmatrix} \boldsymbol{g}^A & \boldsymbol{g}^{AB} \\ \boldsymbol{g}^{BA} & \boldsymbol{g}^B \end{bmatrix}$, and $|\mathbf{F}(\omega)\rangle = \begin{bmatrix} \mathbf{F}^A \\ \mathbf{F}^B \end{bmatrix}$. $\tilde{\boldsymbol{\omega}}^A$ and $\tilde{\boldsymbol{\omega}}^B$ are diagonal matrices containing the complex eigenfrequencies of the uncoupled QNMs. The matrices, $\boldsymbol{\kappa}^{AB}$, $\boldsymbol{\kappa}^{BA}$, $\boldsymbol{g}^{AB}$, and $\boldsymbol{g}^{BA}$ describe the coupling between the resonator QNMs, $\boldsymbol{g}^A$ and $\boldsymbol{g}^B$ represent the self-coupling corrections [35], and $\mathbf{F}^A$ and $\mathbf{F}^B$ account for the coupling between the incident field and the modes of the individual resonators. The explicit expressions of these quantities are

$$\begin{aligned}\kappa_{mn}^{AB} &= \tilde{\omega}_n^B \iiint_\Omega (\varepsilon_\infty^B \tilde{\mathbf{E}}_m^A \cdot \tilde{\mathbf{E}}_n^B - \mu_0 \tilde{\mathbf{H}}_m^A \cdot \tilde{\mathbf{H}}_n^B) d^3\mathbf{r} \\ &+ \tilde{\omega}_n^B \iiint_{\Omega_c} \frac{1}{\varepsilon_\infty}\left(\frac{1}{\omega_p}\right)^2 (\omega_0^2 + \tilde{\omega}_m^A\tilde{\omega}_n^B)\varepsilon_L(\tilde{\omega}_m^A)\varepsilon_L(\tilde{\omega}_n^B)\tilde{\mathbf{E}}_m^A \cdot \tilde{\mathbf{E}}_n^B d^3\mathbf{r} \\ &- \iiint_{\Omega_r^A} \tilde{\omega}_m^A \varepsilon_L^A(\tilde{\omega}_m^A)\tilde{\mathbf{E}}_m^A \cdot \tilde{\mathbf{E}}_n^B\, d^3\mathbf{r},\end{aligned} \tag{4}$$

$$\begin{aligned}g_{mn}^{AB} &= \iiint_\Omega (\varepsilon_\infty^B \tilde{\mathbf{E}}_m^A \cdot \tilde{\mathbf{E}}_n^B - \mu_0 \tilde{\mathbf{H}}_m^A \cdot \tilde{\mathbf{H}}_n^B) d^3\mathbf{r} \\ &+ \iiint_{\Omega_c} \frac{1}{\varepsilon_\infty}\left(\frac{1}{\omega_p}\right)^2 (\omega_0^2 + \tilde{\omega}_m^A\tilde{\omega}_n^B)\varepsilon_L(\tilde{\omega}_m^A)\varepsilon_L(\tilde{\omega}_n^B)\tilde{\mathbf{E}}_m^A \cdot \tilde{\mathbf{E}}_n^B d^3\mathbf{r} \\ &+ \iiint_{\Omega_r^A} \Delta\varepsilon_\infty^A \tilde{\mathbf{E}}_m^A \cdot \tilde{\mathbf{E}}_n^B d^3\mathbf{r},\end{aligned} \tag{5}$$

$$\begin{aligned}g_{mn}^{A} &= \iiint_{\Omega_r^B} (\varepsilon_\infty^B - \varepsilon_b) \tilde{\mathbf{E}}_m^A \cdot \tilde{\mathbf{E}}_n^A d^3\mathbf{r} \\ &+ \delta_{mn},\end{aligned} \tag{6}$$

$$\begin{aligned}\mathrm{F}_m^A &= \iiint_{\Omega_r^B} \omega(\varepsilon_\infty^B - \varepsilon_b) \tilde{\mathbf{E}}_m^A \cdot \mathbf{E}_D d^3\mathbf{r} \\ &+ \iiint_{\Omega_r^A} \tilde{\omega}_m^A \varepsilon_L^A(\tilde{\omega}_m^A)\tilde{\mathbf{E}}_m^A \cdot \mathbf{E}_D d^3\mathbf{r} + \iiint_{\Omega_r^A} \omega(\varepsilon_\infty^A - \varepsilon_b)\tilde{\mathbf{E}}_m^A \cdot \mathbf{E}_D d^3\mathbf{r}.\end{aligned} \tag{7}$$

The definitions of the spatial domains, $\Omega$, $\Omega_r^A$, $\Omega_r^B$, and $\Omega_c$, are illustrated in Figure 1. Here, $\tilde{\mathbf{E}}_m^X$ and $\tilde{\mathbf{H}}_m^X$ denote the normalized electric and magnetic fields of QNM or PML mode $m$ with eigenfrequency $\tilde{\omega}_m^X$. The modes are normalized such that $\iiint_\Omega \left\{\tilde{\mathbf{E}}_m^X \cdot \frac{\partial[\omega\varepsilon(\omega)]}{\partial\omega}\Big|_{\omega=\tilde{\omega}_m^X} \tilde{\mathbf{E}}_m^X - \tilde{\mathbf{H}}_m^X \cdot \mu_0 \tilde{\mathbf{H}}_m^X\right\} d^3\mathbf{r} = 1$, which can also be expressed in the compact form shown in eq S1.11 for $m = n$. $\varepsilon^X$ denotes the permittivity of resonator $X$; it may correspond either to a non-dispersive material or to a dispersive material described by the Lorentz model introduced above.

The quantity, $\varepsilon_\infty^X$ is the permittivity at the infinite frequency, while $\varepsilon_L^X = \varepsilon^X - \varepsilon_\infty^X$. $\omega_p$ and $\omega_0$ are the plasma frequency and resonance frequency in the Lorentz model, respectively. Note that within the region $\Omega_c = \Omega - \Omega_r^A - \Omega_r^B$, there is $\varepsilon^A = \varepsilon^B = \varepsilon$, i.e., the parameters of the Lorentz model are identical. The quantity $\varepsilon_b$ denotes the permittivity of the background medium, corresponding to the system without the two resonators. $\mu_0$ is the permeability in vacuum.

The derivation of eqs. 3-7 is provided in Supporting Information, S2. Briefly, eq 3 is derived by substituting the QNM expansion eq 2 into the governing Maxwell's equation 1, and projecting the resulting equation onto the same QNM basis ($\boldsymbol{\Psi}_m^A$ and $\boldsymbol{\Psi}_m^B$) through overlap integrals over the entire computational domain. In particular, the basis for the projection procedure is not unique, i.e., any linearly independent set of fields instead of $\boldsymbol{\Psi}_m^A$ and $\boldsymbol{\Psi}_m^B$ can be used as the projection basis to form a set of independent linear equations which can be used to determine $\alpha_m^A$ and $\alpha_m^B$. The present choice of the projection basis is mathematically equivalent to the Rayleigh–Ritz method [31], a standard approach for solving partial differential equations (see Supporting Information, S3).

Equations 3-7 constitute the main results of the present work. They have a structure similar to that of the coupled-QNM (cQNM) formalism developed in [23] which relies on the SCAdom formulation. The present EXTdom formulation, however, introduces additional contributions to the coupling coefficients. In particular, the coefficients $\kappa_{mn}^{AB}$ and $g_{mn}^{AB}$ are not solely determined by overlap integrals within the resonators: like in [18], they also involve contributions from integrals over the entire computational domain $\Omega$ and the coupling region $\Omega_c$. When the medium in $\Omega_c$ is nondispersive, the overlap integrals performed over $\Omega_c$ vanish, so that $\kappa_{mn}^{AB}$ and $g_{mn}^{AB}$ only retain the first and third terms. The integrals performed over $\Omega$ or $\Omega_c$ in $\kappa_{mn}^{AB}$ and $g_{mn}^{AB}$ are closely related to the orthogonality properties of QNMs. Indeed, for nondispersive resonators, replacing $[\tilde{\mathbf{E}}_m^A, \tilde{\mathbf{H}}_m^A]$ in the first two terms of $\kappa_{mn}^{AB}$ (or $g_{mn}^{AB}$) by $[\tilde{\mathbf{E}}_m^B, \tilde{\mathbf{H}}_m^B]$ makes the sum of these two integrals equal to the overlap integral involved in the orthogonality and normalization relations between two modes of the same resonator. These contributions can therefore become significant when $\tilde{\mathbf{E}}_m^A$ and $\tilde{\mathbf{E}}_m^B$ have similar spatial distributions, but tend to vanish when the two modes are nearly orthogonal. This feature distinguishes the EXTdom formulation from the SCAdom formulation. In particular, it can lead to coupling coefficients between two QNMs that are substantially larger than those between a QNM and a PML mode. The latter are generally weaker because PML modes are primarily confined to the PML region and consequently have limited spatial overlap with the QNMs of either resonator.

The present formalism retains an important advantage of our previous SCAdom formulation [23], which is not shared by other approaches [21,22]: the matrices $\hat{\mathbf{H}}$ and $\hat{\mathbf{D}}$ are frequency-independent. Consequently, the coupled-QNM problem can be expressed as a linear eigenvalue problem,

$$\hat{\mathbf{H}}|\alpha_m^S\rangle = \tilde{\omega}_m^S \hat{\mathbf{D}}|\alpha_m^S\rangle. \tag{8}$$

The linearity is ensured by the introduction of the auxiliary fields, as discussed above. This formulation offers two major advantages. First, the supermodes can be computed in an ultra-efficient manner. Second, the state vector $|\alpha(\omega)\rangle$ in eq 2 can be readily projected onto the basis formed by the supermodes: $|\alpha(\omega)\rangle = \sum_m c_m^S(\omega)|\alpha_m^S\rangle$ with

$$c_m^S(\omega) = \frac{1}{\tilde{\omega}_m^S - \omega} \frac{\langle \alpha_m^S | \mathbf{F}(\omega) \rangle}{\langle \alpha_m^S | \hat{\mathbf{D}} | \alpha_m^S \rangle}, \tag{9}$$

where $\langle \alpha_m^S|$ is the left eigenvector determined by solving the eigenvalue problem $\langle \alpha_m^S|\hat{\mathbf{H}} = \tilde{\omega}_m^S \langle \alpha_m^S|\hat{\mathbf{D}}$. Note that because $\hat{\mathbf{H}} \neq \hat{\mathbf{H}}^T$ in general, $\langle \alpha_m^S|$ and $|\alpha_m^S\rangle$ are not colinear.

Another interesting observation concerns $g_{mn}^A$. In the cQNM formalism based on SCAdom, $g_{mn}^A$ is simply equal to $\delta_{mn}$. In contrast, the present formulation introduces an additional term that depends on the overlap integrals between QNMs of the same resonator. This additional contribution describes the coupling between different modes within the same cavity, i.e., the self-coupling effect. It arises because the QNMs of an individual cavity no longer satisfy the original biorthogonality relation after the permittivity distribution is modified by the presence of the other resonator [35]. Interestingly, for non-dispersive resonators with $\varepsilon^X = \varepsilon_\infty^X$, by neglecting the inter-cavity coupling terms $\kappa_{mn}^{AB}$, $g_{mn}^{AB}$ as well as the source terms $\mathrm{F}_m^A$, the resonance shift caused by the self-coupling term is the same as that computed with the QNM perturbation formalism [36].

It is also worth discussing the relationship between the present framework and earlier cQNM frameworks in which the coupling coefficients are also evaluated over an extended domain. In [18], a rigorous cQNM formalism was established for nondispersive resonators, leading to a formulation closely analogous to eq 3. The main differences are that the frequency $\omega$ appearing in eq 4 is replaced by $\omega^2/2$ in [18], and that the elements of the matrices $\widehat{\mathbf{H}}$ and $\widehat{\mathbf{D}}$ do not depend on the magnetic fields of the QNMs. These differences stem from the distinct starting points of the two derivations. The formalism in [18] is derived by substituting the field expansion into the Helmholtz equation, in which the frequency appears quadratically, whereas eq 4 follows directly from Maxwell's two curl equations, where the frequency enters linearly.

More generally, different cQNM formulations can be constructed by reformulating Maxwell's equations or by introducing auxiliary fields to linearize their frequency dependence [17,37]. Thus, the resulting coupling matrices and eigenvalue problems depend, to some extent, on the chosen mathematical representation of Maxwell's equations.

The present work builds on the framework of [23] but leads to a different modal representation of the same Maxwell problem. Both formulations are exact in the limit of a complete QNM basis and therefore yield the same superQNM spectrum and electromagnetic response. They nevertheless differ in the definition of their coupling coefficients and, more fundamentally, in how the electromagnetic information carried by the QNM fields is distributed among the modes of the expansion.

This difference originates from the treatment of the scattered fields entering the coupling terms. In [23], which is based on the multiple-scattering formalism, the field scattered by resonator *A* acts as the excitation of resonator *B*. The resulting coupling integrals are therefore restricted to the volume of resonator *B*. In the present formulation, derived through an expansion-and-projection procedure equivalent to the Rayleigh–Ritz method [31], the scattered field of the coupled system is projected onto the QNM basis of each individual resonator through overlap integrals evaluated over the entire computational domain (i.e., the extended domain). This leads to coupling integrals extending beyond the resonator volumes. Although the two constructions become equivalent for a complete modal basis, their truncated representations exhibit markedly different convergence properties.

The SCAdom formulation of [23] has the advantage of providing relatively simple and physically intuitive coupling coefficients, expressed through integrals restricted to the individual resonator volumes. By contrast, the present EXTdom formulation involves additional contributions that make the coupling coefficients more complex and less straightforward to interpret physically. Nevertheless, as demonstrated below, this formulation exhibits substantially better accuracy when a small number of QNMs are retained in the expansions.

## 3. Convergence test

In this section, we assess the convergence of the present method in predicting the eigenfrequencies and optical response of coupled dispersive resonators. As illustrated in Figure 2(a), we consider a dimer consisting of two identical silicon two-dimensional (2D) nanowires illuminated under transverse electric (TE) polarization. The permittivity of silicon is described by a single-pole Lorentz model. This configuration is identical to that investigated in our previous work [23], allowing a direct comparison of the convergence properties of the cQNM frameworks based on the present EXTdom formulation and the SCAdom formulation.

Despite its seemingly simple geometry, this system provides a particularly stringent test of convergence. At visible frequencies, the dominant QNMs are weakly confined, resulting in strong radiation losses and quality factors below 5. In our previous work [23], achieving an accurate prediction of the eigenfrequencies, with relative errors below 1%, required approximately 100 uncoupled QNMs and PML modes of the individual resonators.

We first compute the QNMs and PML modes of the individual silicon nanowire using the QNM solver QNMEig implemented in the MAN freeware [38]. Since QNMs and PML modes are treated equivalently within the cQNM framework, we do not distinguish between them in the following convergence analysis. The supermodes of the coupled system are then obtained by solving the eigenvalue problem of eq 8. Specifically, we first solve eq 8 using a sufficiently large set of basis, where the truncation ranks of the individual resonators are given by $N_A = N_B = 412$. The fields of the resulting supermodes can then be expressed as superpositions of the modes of the individual resonators: $\widetilde{\mathbf{\Psi}}_n^S(\mathbf{r},\omega) = \sum_{X=A,B;m=1,2,\ldots N_X} \tilde{\alpha}_m^X(\omega)\widetilde{\mathbf{\Psi}}_m^X(\mathbf{r})$. Note that the modes appearing in the sum are ranked according to their values of $|\tilde{\alpha}_m^X(\omega)|$.

The left panel in Figure 2(b) shows the convergence of the predictions obtained with the present cQNM framework. Specifically, we compare the eigenfrequencies $\tilde{\omega}_n^S(cQNM)$ obtained from the cQNM formalism with the reference value $\tilde{\omega}_n^S(exact)$, calculated directly for the dimer system using the MAN freeware [38]. The relative error $|\tilde{\omega}_n^S(cQNM) - \tilde{\omega}_n^S(exact)|/|\tilde{\omega}_n^S(exact)|$ is plotted as a function of the number of modes retained from the uncoupled resonators.

Three modes with different quality factors are considered. The mode with a relatively high quality factor ($Q = \mathrm{Re}(\tilde{\omega}_m^S)/[-2\mathrm{Im}(\tilde{\omega}_m^S)]$=43) exhibits the fastest convergence. By including only one QNM, a relative error below 1% can already be achieved. For the low-$Q$ mode ($Q$=1.4) and the intermediate-$Q$ mode ($Q$=14) with high radiative loss, an error below 1% is obtained when 10 modes from each individual resonator are included. Remarkably, this represents a significant improvement in accuracy compared with the cQNM framework derived from the SCAdom approach [23]. The convergence behavior of the latter is shown in the right panel of Figure 2(b). For the low- and intermediate-$Q$ modes, accurate predictions can only be obtained when more than 100 modes are included. Note that for the high-$Q$ mode, both methods achieve high accuracy, with errors below 1%, when only a single QNM is considered. The improved convergence of the present method will be discussed in more detail in later sections.

Finally, we assess the capability of the present method to reconstruct the optical response of the coupled system. The extinction spectra of the dimer under normal-incidence plane-wave illumination are calculated by including different numbers of modes. The spectra are obtained by calculating eq 9 and subsequently reconstructing the electric field inside the resonators using eq 2, see Supporting Information, S4 for the details. The results are compared with those from direct real-frequency simulations performed using COMSOL. As the number of included modes increases, the spectra obtained from the cQNM formalism progressively converge toward the exact results. Excellent agreement is achieved when 50 modes of each individual resonator are included. The small discrepancy at shorter wavelengths mainly originates from the neglect of higher-order QNMs of the uncoupled resonators. By further increasing the number of included modes to 400, nearly perfect agreement is obtained, demonstrating the convergence of the present method in predicting the optical response.

The convergence tests have validated the cQNM methods derived using both approaches. However, the true advantage of the theory lies in its ability to accurately predict the eigenfrequencies and optical responses of coupled systems using only a small number of QNMs with clear physical significance. In our previous work, we demonstrated that the cQNM theory based on the SCAdom can provide highly accurate predictions using a limited number of QNMs only for resonators with strong photon confinement, such as subwavelength plasmonic antennas and photonic microcavities. For optical resonators with high radiative losses or weak field confinement, the accuracy deteriorates significantly. In the following section, we show that the cQNM method based on the EXTdom approach can substantially improve the prediction accuracy when only a small number of QNMs are considered.

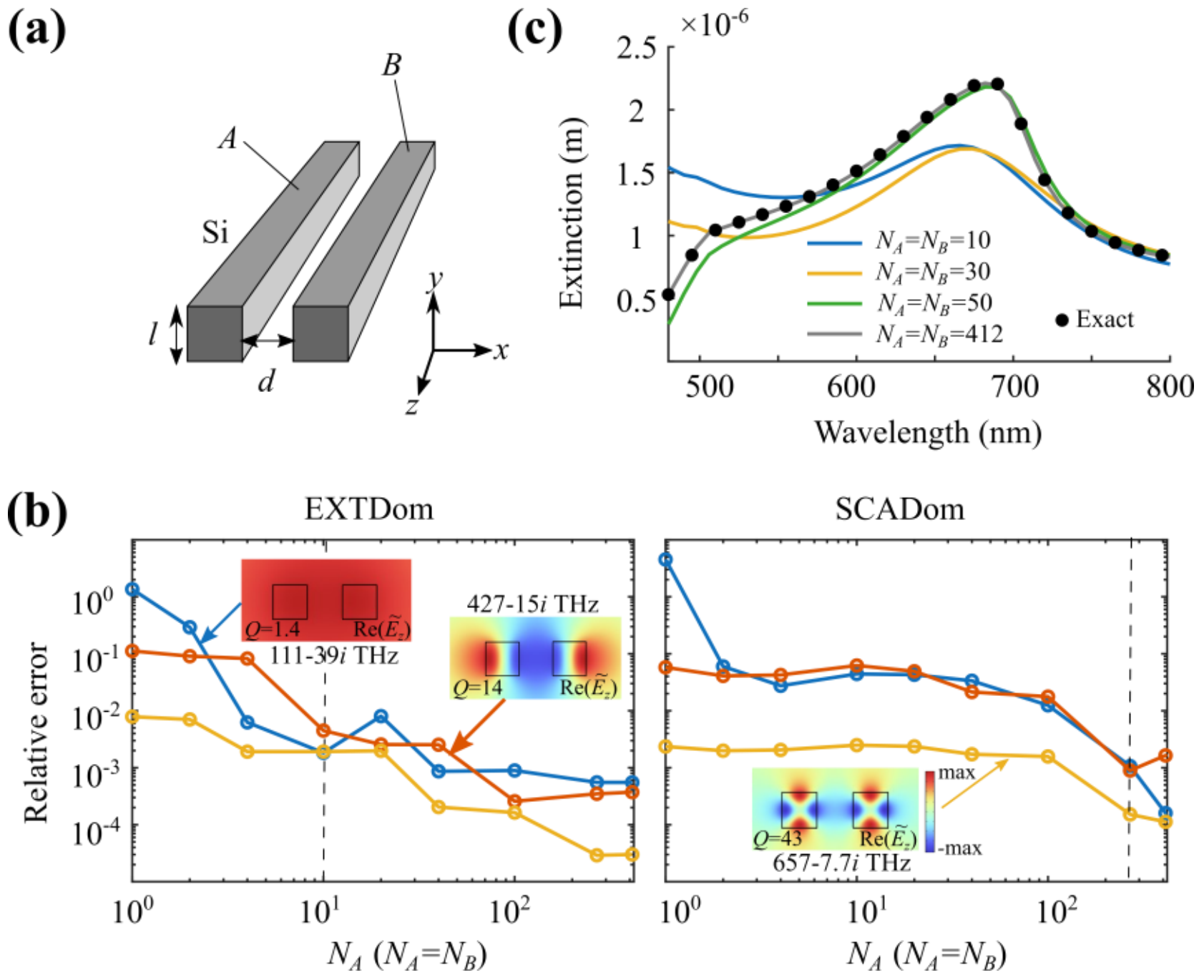

**Figure 2 Convergence of the cQNM framework for a dispersive dimer.** (**a**) Geometry of two identical infinitely long Si nanowires with side length $l$ = 114 nm and separation distance $d$ =116 nm. (**b**) Eigenfrequency convergence for three supermodes (insets). The error is defined as $|\tilde{\omega}_n^S(cQNM) - \tilde{\omega}_n^S(exact)|/|\tilde{\omega}_n^S(exact)|$. Left: present EXTdom framework; right: previous SCA-dom framework [23]. Dashed lines indicate the number of QNMs required to achieve a relative error below $10^{-2}$. (**c**) Convergence of the extinction cross-section spectra calculated using the cQNM theory with $N_A = N_B$ =10, 30, 50, and 412 supermodes, compared with the results of full-wave COMSOL simulations at real frequencies ("Exact"). Silicon dispersion is described by a single-pole Lorentz model with $\varepsilon_\infty$ =8.51$\varepsilon_0$, $\omega_p = 3.62 \times 10^{15}$ rad· s$^{-1}$, $\omega_0 = 5.09 \times 10^{15}$ rad· s$^{-1}$, and $\gamma = 1.16 \times 10^{14}$ rad· s$^{-1}$.

## 4. Accurate prediction with few uncoupled QNMs

In this section, we demonstrate that the present EXTdom cQNM framework enables accurate predictions of eigenfrequencies and optical responses when only a small number of QNMs of the individual resonators are considered. It can enable accurate prediction of the eigenfrequencies and optical responses for optical resonators with high radiative losses or weak field confinement, which is difficult for the cQNM theory based on SCAdom formulation.

### 4.1 Accurate prediction of the eigenfrequency

In this section, we show that the present cQNM framework can accurately predict the eigenfrequencies of supermodes using only a small set of QNMs of the uncoupled resonators.

We consider the same structure as in [18], consisting of a dimer formed by two non-dispersive 2D nanowires with permittivity $\varepsilon = 16$, embedded in a vacuum. As shown in Figure 3(a), the nanowires have a rectangular cross section with widths $L_x = 50$ nm and $L_y = 150$ nm. The structure supports QNMs with low confinement of field, which serves as an excellent example to demonstrate the advantage of the present method.

We first use the MAN freeware to compute two QNMs of the individual resonators, whose field distributions shown in the last panel of Figure 3(b) correspond to dipolar and monopolar modes. We consider the basis constituting these two QNMs, $(\tilde{\mathbf{E}}_m, \tilde{\mathbf{H}}_m, \tilde{\omega}_m)$ with $m = 1, 2$, augmented by the two associated QNMs $(\tilde{\mathbf{E}}_m, -\tilde{\mathbf{H}}_m, -\tilde{\omega}_m)$ with negative frequencies. Note that including the negative-frequency QNMs is essential for accurately predicting the eigenfrequencies of the coupled system. This is because positive- and negative-frequency QNMs share the same electric-field distribution, resulting in large contributions from the first two terms in eqs. 4 and 5, as we have discussed in section 2. Neglecting negative-frequency modes leads to a significant increase in the prediction error.

Four supermodes with positive real parts of the eigenfrequency are obtained, and their field distributions are shown in Figure 3(b). The supermodes in the first column originate from the coupling between the monopolar QNMs, whereas the other two result from the coupling between the dipolar QNMs.

The solid curves in Figure 3(c1) show the eigenfrequencies of the four supermodes as a function of the gap distance between the nanowires. The exact results, obtained by directly calculating the QNMs of the coupled dimer, are shown with circles. Excellent agreement is achieved for most gap distances. However, when the gap distance decreases below 60 nm, a significant deviation appears in the imaginary part of the second supermode. At small separations, the interaction between the monopolar QNMs becomes extremely strong, and their coupling is mediated by additional PML modes.

The results obtained using the cQNM framework based on the SCAdom are shown in Figure 3(c2). A significant deviation from the exact results can be observed for the imaginary parts of both antisymmetric supermodes.

Figure 3(d) compares the relative errors obtained with the two approaches. Except for the symmetric supermode formed by the coupling of the monopolar modes, the accuracy of all other three supermodes is significantly improved by the present framework. In particular, for the first antisymmetric supermode, the relative error obtained with the SCAdom cQNM framework exceeds 10%, whereas the present framework reduces the error by approximately one order of magnitude.

At present, we cannot provide a rigorous mathematical explanation for this improvement. A possible interpretation is that the coupling coefficients in the EXTdom formulation contain overlap integrals involving the fields outside the resonators, including the PML regions, and therefore provide a more accurate description of the interaction between QNMs with strong radiation leakage or weak confinement. This interpretation is supported by the observation that both methods provide reasonable predictions for the supermodes formed by the coupling of two dipolar modes, which exhibit stronger field confinement. In contrast, the monopolar modes of the individual resonators have weaker confinement and significant field amplitudes extending into the surrounding region and PML layers. Therefore, the eigenfrequencies of the supermodes generated by their coupling can only be accurately predicted by the EXTdom cQNM framework.

Another possible explanation is that, in the cQNM formulation based on SCAdom, an accurate description of the coupling requires an accurate reconstruction of the scattered fields of both resonators, which mutually serve as the incident fields for each other. When the dominant QNMs of the individual resonators exhibit low field confinement, their scattered fields contain substantial contributions from non-resonant components, such that an accurate reconstruction may require a large number of non-resonant QNMs and PML modes. In contrast, the cQNM formulation based on EXTdom avoids the intermediate reconstruction of the scattered field of each individual resonator and instead directly projects the Maxwell's equations of the coupled system onto the basis of individual-resonator QNMs. This provides a more direct representation of the coupled system eigenfields and can therefore achieve faster convergence with respect to the number of retained modes.

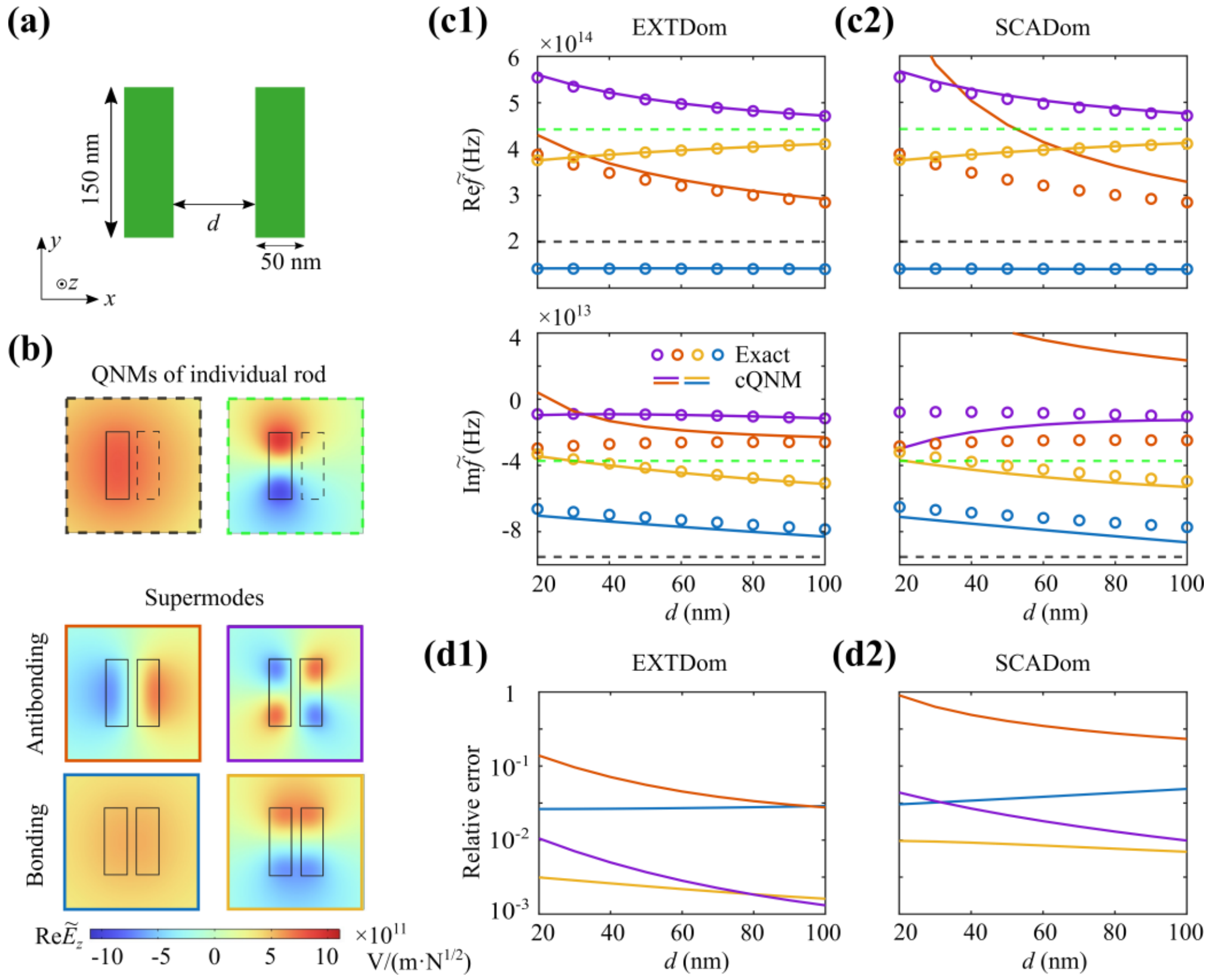

**Figure 3 Accurate prediction of supermode eigenfrequencies using a small set of QNMs of the uncoupled resonators.** (**a**) Schematic of the 2D dimer. TE-polarized QNMs are considered. (**b**) Top: normalized *z*-component of the electric field ($\mathrm{Re}(\tilde{E}_z^A)$) of the monopolar and dipolar QNMs of the left resonator (solid rectangle). Bottom: electric-field distributions of the corresponding supermodes obtained by coupling the monopolar or dipolar QNMs of the two resonators. Field maps are shown for a gap of *d* = 20 nm; eigenfrequencies are given in (**c**). (**c**) Real and imaginary parts of the eigenfrequencies as a function of the gap distance. The cQNM predictions (solid curves) are compared with the reference results (circles) obtained by directly computing the QNMs of the coupled structure with MAN. The dashed lines indicate the eigenfrequencies of the individual resonator. (**d**) Relative errors, $|\tilde{\omega}_n^S(cQNM) - \tilde{\omega}_n^S(exact)|/|\tilde{\omega}_n^S(exact)|$, of the predicted eigenfrequencies. Panels (**c1,d1**) and (**c2,d2**) show results from the EXTdom and SCAdom frameworks, respectively.

## 4.2 Accurate prediction of the optical response

In this section, we show that the present cQNM framework also provides improved accuracy in reconstructing the optical response of coupled resonators. To illustrate this capability, we consider plasmonic molecules composed of two identical nanorods. As shown in Figure 4, three dimers are considered. The three systems have the same geometry, consisting of two parallel nanorods with the same aspect ratio, while the ratio between the rod gap and rod length is kept fixed. They differ only in their overall size. We calculate the extinction spectra of these three systems under illumination by a plane wave incident at an angle of $\theta$=55° (between the wavevector $\mathbf{k}$ and the normal vector of the plane determined by the two parallel long axes of the nanorods), with the electric field $\mathbf{E}_{inc}$ polarized along the long axis of the nanorods.

We expect that the optical response becomes increasingly difficult to reconstruct if using a small set of QNMs as the size of the system increases. This can be understood from the changing field confinement. For small resonators, the electric field is strongly confined to the immediate vicinity of the resonator, and the QNMs of the individual resonators therefore provide the dominant contribution to the optical response, while the contribution of PML modes remains relatively small. Neglecting these PML modes consequently introduces only a minor error. As the size of the resonators increases, however, the field confinement becomes weaker, and a larger fraction of the field extends into the surrounding region. As a result, the contribution of PML modes becomes increasingly important and can no longer be neglected. In this section, we demonstrate that the present cQNM framework is particularly advantageous in addressing this challenge for large resonators.

Similar to the previous section, we evaluate the accuracy of the method when only a small set of QNMs of the individual resonators is considered. Here, we include two dipolar QNMs for each nanorod: one with a positive eigenfrequency and the other with a negative eigenfrequency. Unlike the non-dispersive case, however, the negative-frequency dipolar QNM does not have an eigenfrequency that is exactly opposite to that of the positive-frequency mode and therefore has to be computed independently.

Using the eigenfrequencies and QNM fields of the uncoupled nanorods, we construct the 4×4 coupling matrices $\widehat{\mathbf{H}}$ and $\widehat{\mathbf{D}}$. Solving the homogeneous eq 8 then yields the eigenfrequencies and eigenvectors of the superQNMs. The total extinction cross section is subsequently reconstructed by summing the contributions from all superQNMs using eqs. 2 and 9.

The extinction spectra obtained with the present EXTdom framework are shown in the first column of Figure 4. The reconstructed extinction spectra (black solid curves) exhibit nice agreement with the exact results obtained from real-frequency simulations using COMSOL (black circles). In addition to the total extinction, we also calculate the modal contributions to the extinction from two supermodes with $\mathrm{Re}(\widetilde{\omega}_m) > 0$. We also checked these results by comparing them with those calculated directly from the exact superQNMs. Excellent agreement is observed between the two approaches (not shown here for conciseness). Notably, the discrepancy between the reconstructed spectra and the real-frequency results for $s = 2$ in the shorter-wavelength regime does not reflect an inaccuracy of the EXTdom framework, but rather results from the omission of higher-order supermodes.

The corresponding results obtained using the cQNM framework based on SCAdom are shown in the second column of Figure 4 for comparison. For the dimer with the smallest size, the SCAdom cQNM framework also provides a good agreement with the real-frequency simulation. However, as the size of the dimer increases, increasingly pronounced discrepancies appear in both the amplitudes and spectral positions of the resonance peaks. In contrast, the EXTdom cQNM framework maintains a high level of accuracy for all three systems, demonstrating its improved capability to reconstruct the optical response of large resonators using only a small set of physically meaningful QNMs.

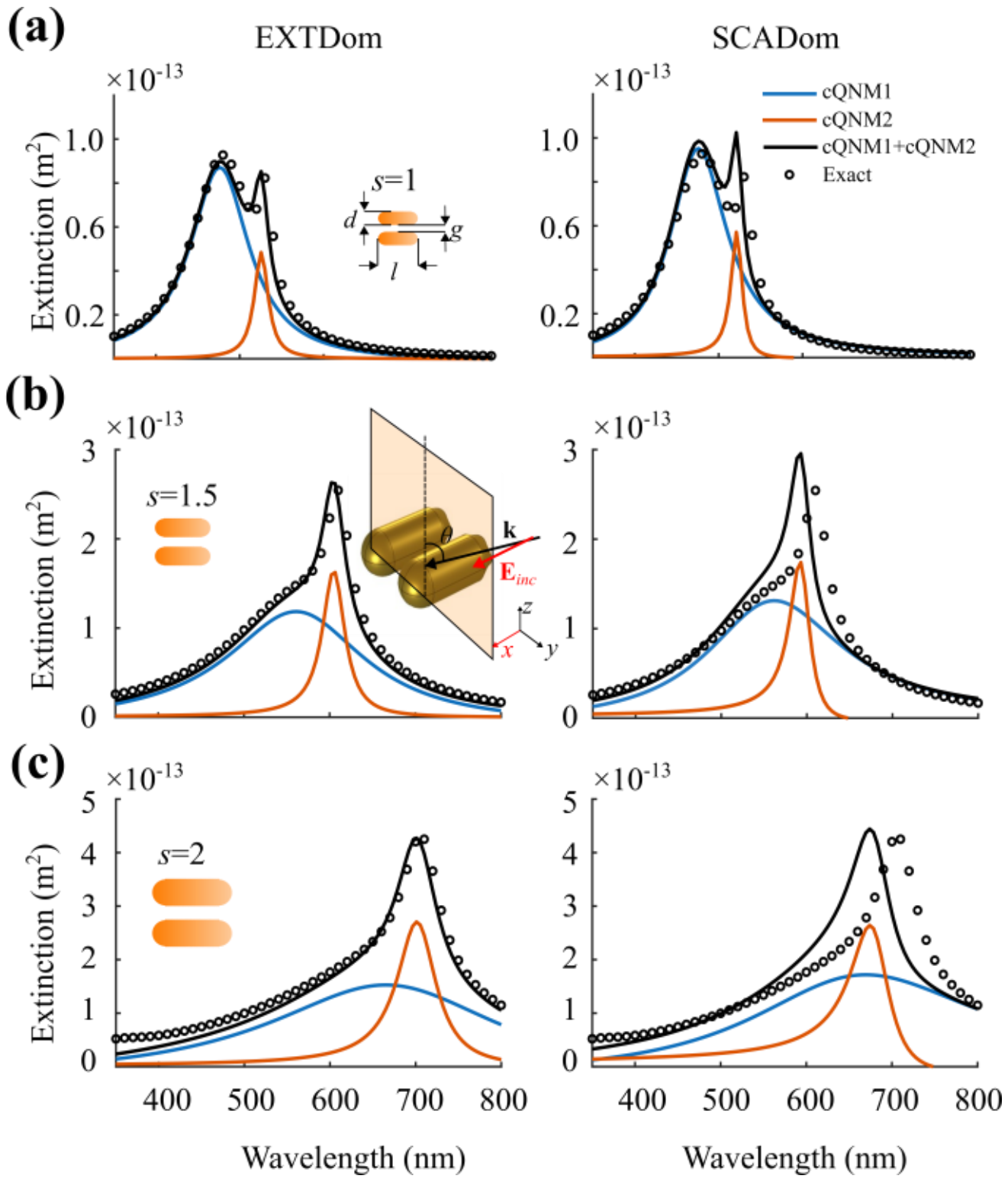

**Figure 4 Calculation of the extinction spectra of Au nanorod dimers with different sizes.** The system consists of two identical Au nanorods embedded in air, each with a length $l = s \times 100$ nm and a diameter of $d = s \times 40$ nm, separated by a gap of $g = s \times 20$ nm. (**a**), (**b**), and (**c**) correspond to dimers with $s = 1$, 1.5, and 2, respectively. The frequency-dependent permittivity of gold is described by a Drude model, $\varepsilon(\omega) = \varepsilon_\infty - \varepsilon_\infty \frac{\omega_p^2}{\omega^2 + i\omega\gamma}$, with $\varepsilon_\infty = \varepsilon_0$, $\omega_p = 1.26 \times 10^{16}$ rad$\cdot$ s$^{-1}$, and $\gamma = 1.41 \times 10^{14}$ rad$\cdot$ s$^{-1}$. The dimers are excited by a plane wave incident at an angle of $\theta = 55°$ with the electric field $\mathbf{E}_{inc}$ parallel to the long axis of the nanorods. The extinction spectra calculated using the cQNM frameworks (black solid curves) are compared with the "Exact" results (black circles) obtained from real-frequency simulations using COMSOL. The contributions of the two supermodes are shown by the blue and orange curves. The left and right panels correspond to the EXTdom and SCAdom frameworks, respectively.

## 5. Practical method for computing the coupling coefficients

In the previous section, we showed that the cQNM framework based on EXTdom provides significantly improved accuracy in predicting the responses of coupled systems using only a small set of individual QNMs. However, the coupling coefficients $\kappa_{mn}^{AB}$ and $g_{mn}^{AB}$ in the present formulation involve integrals evaluated over the entire computational domain $\Omega$, making their computation more expensive than that of the corresponding quantities in our previous work based on SCAdom [23]. Moreover, evaluating these integrals requires the computational domains of the two coupled resonators, including both the physical and PML regions, to overlap. This requirement leads to computational difficulties when the relative positions of the resonators are varied during the design of a coupled system. In this section, we overcome these limitations by reformulating the coupling coefficients in terms of simpler integrals evaluated only on or within the resonator boundaries.

Let us first consider the coupling coefficient $\kappa_{mn}^{AB}$. Our goal is to reformulate the first and second terms in eq 4 as integrals evaluated only within or on the surfaces of the two coupled resonators. To this end, we decompose the first term into two integrals over the regions $\Omega_r^A \cup \Omega_r^B$ and $\Omega_c$, respectively, and combine the latter with the second term in eq 4. This yields a term $S$, given by

$$S = \tilde{\omega}_n^B \iiint_{\Omega_c} \left[ \frac{1}{\varepsilon_\infty} \left( \frac{1}{\omega_p} \right)^2 (\omega_0^2 + \tilde{\omega}_m^A \tilde{\omega}_n^B) \varepsilon_L(\tilde{\omega}_m^A) \varepsilon_L(\tilde{\omega}_n^B) \tilde{\mathbf{E}}_m^A \cdot \tilde{\mathbf{E}}_n^B + \varepsilon_\infty \tilde{\mathbf{E}}_m^A \cdot \tilde{\mathbf{E}}_n^B - \mu_0 \tilde{\mathbf{H}}_m^A \cdot \tilde{\mathbf{H}}_n^B \right] d^3\mathbf{r}. \quad (10)$$

This volume integral can be converted to a surface integral using the Lorentz reciprocity theorem [33] and the definition of $\varepsilon_L$. Namely, we find

$$S = i\tilde{\omega}_n^B(\tilde{\omega}_m^A - \tilde{\omega}_n^B)^{-1} \iint_{\Sigma_r^A \cup \Sigma_r^B} (\tilde{\mathbf{E}}_n^B \times \tilde{\mathbf{H}}_m^A - \tilde{\mathbf{E}}_m^A \times \tilde{\mathbf{H}}_n^B) \cdot \mathbf{n}\, d^2\mathbf{r}, \quad (11)$$

$\mathbf{n}$ being the normal outwards-pointing unit vector on the surface of the resonator, $\Sigma_r^A$ or $\Sigma_r^B$. Note that to arrive at eq 11, one should also guarantee that $(\tilde{\mathbf{E}}_n^B \times \tilde{\mathbf{H}}_m^A) \cdot \mathbf{n}$ and $(\tilde{\mathbf{E}}_m^A \times \tilde{\mathbf{H}}_n^B) \cdot \mathbf{n}$ vanish at the boundary of PML. This condition is automatically satisfied for QNMs whose fields decay to zero at the boundary of the PML. For PML modes, it can be ensured by surrounding the computational domain with either perfectly electric conducting (PEC) or perfectly magnetic conducting (PMC) boundaries, a standard practice when solving modes of resonators.

Using eq 11, the coupling coefficient can be reformulated as

$$\kappa_{mn}^{AB} = \tilde{\omega}_n^B \iiint_{\Omega_r^A \cup \Omega_r^B} (\varepsilon_\infty^B \tilde{\mathbf{E}}_m^A \cdot \tilde{\mathbf{E}}_n^B - \mu_0 \tilde{\mathbf{H}}_m^A \cdot \tilde{\mathbf{H}}_n^B) d^3\mathbf{r}$$

$$+ i\tilde{\omega}_n^B(\tilde{\omega}_m^A - \tilde{\omega}_n^B)^{-1} \iint_{\Sigma_r^A \cup \Sigma_r^B} (\tilde{\mathbf{E}}_n^B \times \tilde{\mathbf{H}}_m^A - \tilde{\mathbf{E}}_m^A \times \tilde{\mathbf{H}}_n^B) \cdot \mathbf{n} d^2\mathbf{r}$$

$$- \iiint_{\Omega_r^A} \tilde{\omega}_m^A \varepsilon_L^A(\tilde{\omega}_m^A) \tilde{\mathbf{E}}_m^A \cdot \tilde{\mathbf{E}}_n^B\, d^3\mathbf{r}. \quad (12)$$

Following a similar procedure, the coupling coefficient $g_{mn}^{AB}$ can be expressed as

$$g_{mn}^{AB} = \iiint_{\Omega_r^A \cup \Omega_r^B} (\varepsilon_\infty^B \tilde{\mathbf{E}}_m^A \cdot \tilde{\mathbf{E}}_n^B - \mu_0 \tilde{\mathbf{H}}_m^A \cdot \tilde{\mathbf{H}}_n^B) d^3\mathbf{r}$$

$$+ i(\tilde{\omega}_m^A - \tilde{\omega}_n^B)^{-1} \iint_{\Sigma_r^A \cup \Sigma_r^B} (\tilde{\mathbf{E}}_n^B \times \tilde{\mathbf{H}}_m^A - \tilde{\mathbf{E}}_m^A \times \tilde{\mathbf{H}}_n^B) \cdot \mathbf{n} d^2\mathbf{r}$$

$$+ \iiint_{\Omega_r^A} \Delta\varepsilon_\infty^A \tilde{\mathbf{E}}_m^A \cdot \tilde{\mathbf{E}}_n^B d^3\mathbf{r}. \quad (13)$$

Note that the term $(\tilde{\omega}_m^A - \tilde{\omega}_n^B)^{-1}$ becomes divergent when $\tilde{\omega}_m^A = \tilde{\omega}_n^B$. In this case, the corresponding surface-integral terms should instead be reformulated as:

$$(\tilde{\omega}_m^A - \tilde{\omega}_n^B)^{-1} \iint_{\Sigma_r^A \cup \Sigma_r^B} (\tilde{\mathbf{E}}_n^B \times \tilde{\mathbf{H}}_m^A - \tilde{\mathbf{E}}_m^A \times \tilde{\mathbf{H}}_n^B) \cdot \mathbf{n} d^2\mathbf{r}$$

$$= \left[\frac{\partial \tilde{\omega}_m^A(p)}{\partial p}\right]^{-1} \iint_{\Sigma_r^A \cup \Sigma_r^B} \left[\tilde{\mathbf{E}}_n^B \times \frac{\partial \tilde{\mathbf{H}}_m^A(p)}{\partial p} - \frac{\partial \tilde{\mathbf{E}}_m^A(p)}{\partial p} \times \tilde{\mathbf{H}}_n^B\right] \cdot \mathbf{n}\, d^2\mathbf{r}, \quad (14)$$

where $p = p_0$ and $p$ can be any parameter of system $A$ (for instance, the refraction index or the size) such that at $p = p_0$, the system $A$ becomes identical to the system $B$, thus leading to $\tilde{\omega}_m^A(p) = \tilde{\omega}_n^B$, $\tilde{\mathbf{E}}_m^A(p) = \tilde{\mathbf{E}}_n^B$, and $\tilde{\mathbf{H}}_m^A(p) = \tilde{\mathbf{H}}_n^B$.

In Figure 5, we consider two coupled resonators residing on a dispersive substrate, as illustrated in Figure 5(a). Note that in such a challenging case, the second terms in eq 4 and eq 5 are non-zero, making the computation of $\kappa_{mn}^{AB}$ and $g_{mn}^{AB}$ quite complicated. By using eq 12 and eq 13, the computation only involves the evaluation of a surface integral, and some volume integrals performed inside the resonator volumes. As demonstrated in Figure 5(b), the coupling coefficients $g_{mn}^{AB}$ for the modes in Figure 5(c), evaluated using eq (13) show a perfect agreement with those calculated with eq 5.

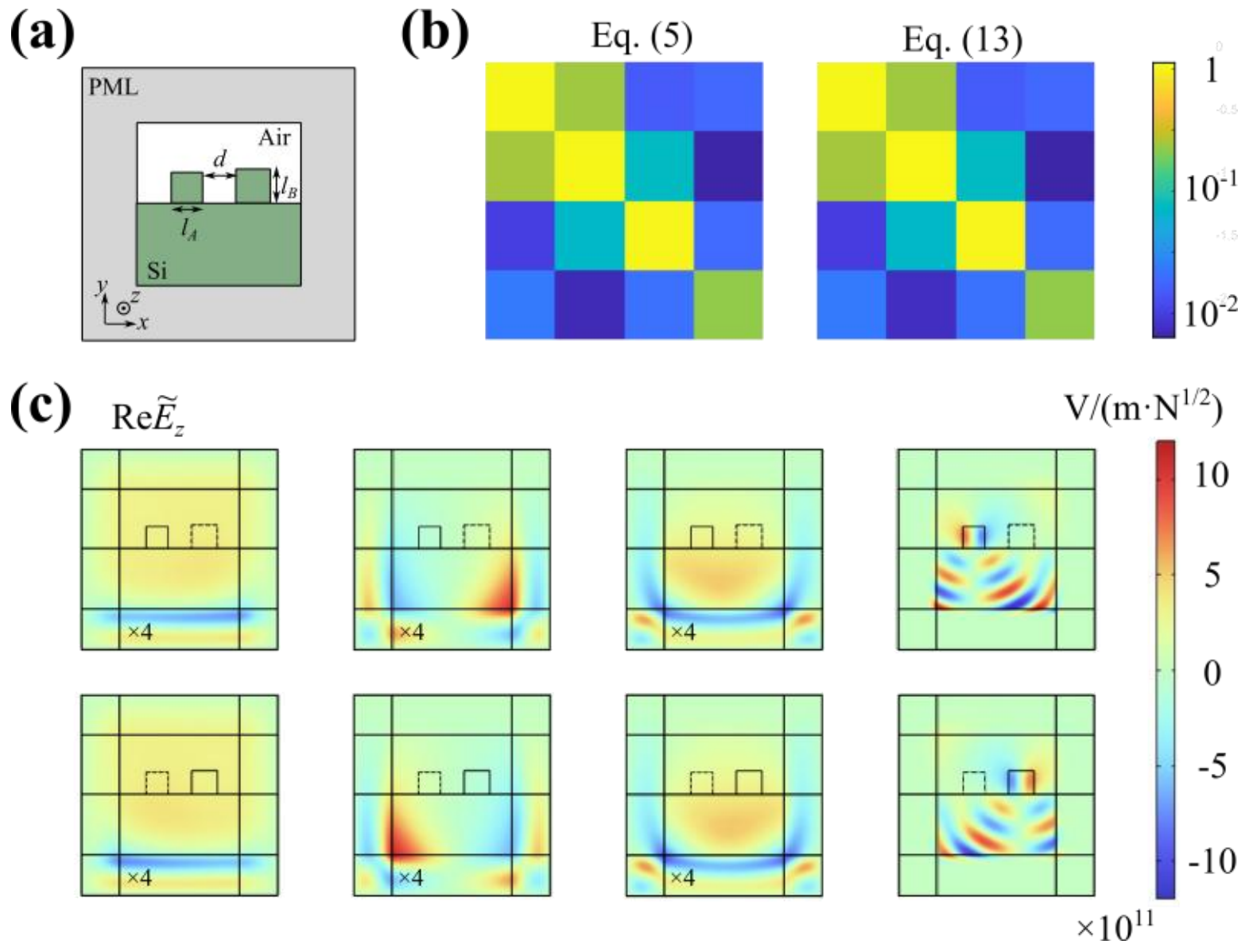

**Figure 5 Efficient computation of the coupling coefficients $\mathbf{g}_{mn}^{AB}$.** (**a**) Geometry of the coupled system, consisting of two Si nanowires with side lengths $l_A$ = 114 nm and $l_B$ = 125 nm, separated by a gap of $d$ = 122 nm. The two-dimensional system is considered with TE polarization. The nanowires are placed in air on a Si substrate, and the frequency-dependent permittivity of Si is described by the same single-pole Lorentz model and parameters as in Figure 2. The whole structure is surrounded by PMLs in $x$- and $y$-directions. (**b**) Absolute values of coupling coefficients $g_{mn}^{AB}$ for four pairs of modes (field distributions shown in (**c**)) computed using eqs. 5 and 13, respectively. (**c**) Normalized $z$-component of the electric field in the $(x, y)$ plane for the four pairs of modes of the individual nanowire (shown by the square solid line). The fields of the first three PML modes are significantly weaker than that of the last QNM and are multiplied by 4 for a clear display.

## 6. Conclusion and perspectives

In summary, we have developed an improved cQNM framework for analyzing dissipative coupling between photonic and plasmonic resonators. Compared with [23], it provides substantially more accurate predictions of coupled resonators using only a small number of QNMs of the individual resonators. This improvement is particularly pronounced for resonators supporting highly radiative, weakly confined QNMs, whose fields extend significantly beyond the resonator volumes. The ability to achieve accurate predictions with compact QNM expansions considerably extends the applicability of cQNM theory to coupled-resonator systems.

An important question, however, is when to use the present framework rather than the previous formulation. The improved convergence comes at the cost of a more involved mathematical formulation. In particular, the coupling coefficients contain additional contributions, making their numerical evaluation more demanding and their physical interpretation less straightforward. In the SCAdom framework [23], the coupling between QNMs can, in many cases, be directly associated with either $\kappa_{mn}^{AB}$ or $g_{mn}^{AB}$. For example, for non-dispersive resonators, $g_{mn}^{AB} \neq 0$ while $\kappa_{mn}^{AB} = 0$. Conversely, in strongly dispersive systems, such as plasmonic resonators, where $g_{mn}^{AB} \neq 0$ and $\kappa_{mn}^{AB} \neq 0$, the coupling is dominated by $\kappa_{mn}^{AB}$ and the contribution of $g_{mn}^{AB}$ can often be neglected.

This dominance by either $\kappa_{mn}^{AB}$ or $g_{mn}^{AB}$ provides a relatively transparent physical interpretation of the coupling mechanisms. In particular, the real and imaginary parts of the coupling coefficients can be associated with coherent and dissipative coupling, respectively, the former generally leading to level repulsion and the latter potentially giving rise to level attraction. Such a direct interpretation is more difficult within the present formulation. Indeed, the first and second integral contributions to $\kappa_{mn}^{AB}$ in eq 4 can be comparable to those contributing to $g_{mn}^{AB}$ in eq 5. Neither coupling coefficient can therefore generally be neglected, making it more difficult to isolate individual coupling mechanism.

Another practical consideration is the inclusion of negative-frequency QNMs. The present framework requires including these modes to accurately predict the supermode eigenfrequencies, whereas they can generally be neglected in the SCAdom formulation [23]. This represents an additional advantage of the latter when sufficiently accurate predictions can be obtained using only a small number of positive-frequency QNMs.

Based on these considerations, we recommend using the cQNM framework in [23] for resonators whose QNM fields are tightly confined within their volumes. In this regime, both formulations can generally provide accurate predictions, while SCAdom offers simpler coupling coefficients and a more transparent physical interpretation. By contrast, the present EXTdom framework is particularly advantageous for coupled resonators supporting low-$Q$ or highly radiative QNMs, whose fields extend substantially beyond the resonator volumes, as is often the case for large plasmonic nanoparticles. For such systems, SCAdom may require a large number of QNMs to achieve accurate predictions, whereas EXTdom can maintain high accuracy with a relatively small modal basis.

Looking ahead, there is a growing interest in the simulation of large-scale array of optical nanoresonators with overall dimensions much larger than the wavelength, for which cQNM-based approaches have recently demonstrated computational speedups of one to two orders of magnitude while maintaining good accuracy [39,40]. To date, these approaches have primarily been applied to arrays of plasmonic resonators with strongly confined fields. Their applicability to systems composed of relatively low-index meta-atoms [28-30] remains largely unexplored. Given its improved convergence for weakly confined QNMs, the present framework could provide a useful basis for extending cQNM-based simulations to metasurfaces and large-scale arrays composed of such meta-atoms. This potential application deserves further investigation.

The codes and COMSOL models used to generate the results presented in this work will be incorporated into the MAN freeware (Version 9.1 and subsequent releases). MAN Version 9.0 [41] already includes a cQNM toolbox based on the framework in [23]. Making both formulations available within the same software environment will allow users to compare their respective performances and select the most appropriate approach for a given coupled-resonator system.

## 7. Acknowledgements

Most of this work was performed during the postdoctoral appointment of TW at LP2N. PL acknowledges funding from the UPiCO project (ANR-25-CE09-4019). HL acknowledges funding from the National Natural Science Foundation of China (62475120, 62535007).

## 8. References

1. H. Benisty, J. J. Greffet, and P. Lalanne, Introduction to Nanophotonics, Oxford University Press, Oxford (2022).
2. N. J. Halas et al., “Plasmons in strongly coupled metallic nanostructures,” Chem. Rev. 111, 3913–3961 (2011)
3. T. Wu and P. Lalanne, “Designing electromagnetic resonators with quasinormal modes,” Front. Phys. 12, 1461106 (2024).
4. D. Song et al., “Enhanced light–matter interaction in plasmonic nanocavities: principle, challenge, and perspective,” Photonics Insights 4, R09 (2026).
5. C. Heimig, E. Baù, J. Biechteler, et al. Metasurfaces for the nanoscale engineering of light–matter coupling. Nat. Rev. Mater. https://doi.org/10.1038/s41578-026-00951-1 (2026).
6. Y. Chen et al., “Observation of intrinsic chiral bound states in the continuum,” Nature 613, 474–478 (2023).
7. H. Deng et al., “Chiral lasing enabled by strong coupling,” Sci. Adv., 11, eads9562 (2025).
8. Y. P. Wang and C. M. Hu, “Dissipative couplings in cavity magnonics,” J. Appl. Phys. 127, 130901 (2020).
9. M. Harder, B. M. Yao, Y. S. Gui, and C. M. Hu, “Coherent and dissipative cavity magnonics,” J. Appl. Phys. 129, 201101 (2020).
10. A. Bogdanov, K. Koshelev, P. Kapitanova, et al., “Bound States in the Continuum and Fano Resonances in the Strong Mode Coupling Regime,” Adv. Photonics 1, 016001 (2019).
11. M. Mikhailovskii, M. Poleva, N. Solodovchenko, et al., “Engineering of high-Q states via collective mode coupling in chains of Mie resonators,” ACS Photon. 11:1657 (2024).
12. C. Wang et al., “Non-Hermitian optics and photonics: from classical to quantum,” Adv. Opt. Photonics 15, 442–523 (2023).
13. A. Li et al., “Exceptional points and non-Hermitian photonics at the nanoscale,” Nat. Nanotechnol. 18, 706–720 (2023).
14. H. A. Haus and W. Huang, “Coupled-mode theory,” Proc. IEEE 79, 1505–1518 (1991).
15. T. Christopoulos, O. Tsilipakos, and E. E. Kriezis, “Temporal coupled-mode theory in nonlinear resonant photonics,” J. Appl. Phys. 136, 011101 (2024).
16. A. Frisk Kockum, A. Miranowicz, S. De Liberato, et al. Ultrastrong coupling between light and matter. Nat. Rev. Phys., 1, 19 (2019).
17. C. Sauvan et al., “Normalization, orthogonality, and completeness of quasinormal modes of open systems: the case of electromagnetism,” Opt. Express 30, 6846–6885 (2022).
18. B. Vial and Y. Hao, “A coupling model for quasi-normal modes of photonic resonators,” J. Opt. 18, 115004 (2016).
19. A. C. Tasolamprou et al., “Toroidal eigenmodes in all-dielectric metamolecules,” Phys. Rev. B 94, 205433 (2016).

20. K. G. Cognée, Hybridization of Open Photonic Resonators, PhD dissertation, University of Bordeaux & Amsterdam, Chap. 3(2020).
21. C. Tao et al., “Coupling theory of quasinormal modes for lossy and dispersive plasmonic nanoresonators,” Phys. Rev. B 102, 045430 (2020).
22. J. Ren, S. Franke, and S. Hughes, “Quasinormal modes, local density of states, and classical Purcell factors for coupled loss-gain resonators,” Phys. Rev. X 11, 041020 (2021).
23. T. Wu and P. Lalanne, “Dissipative coupling in photonic and plasmonic resonators,” Adv. Photonics, 7, 056011 (2025).
24. E. Muljarov, “Rigorous theory of coupled resonators,” Phys. Rev. Lett., 136, 023801 (2026).
25. J. Fischbach, S. Gladyshev, A. C. Valero, et al., “Resonant States Reveal Strong Light–Matter Coupling in Nanophotonic Cavities,” Laser & Photonics Rev., 20, e03157 (2026).
26. C. Tao, J. Zhu, and H. Liu, “Impact of mode completeness on the accuracy of the coupling theory of quasinormal modes: a strict numerical demonstration,” arXiv:2606.11630 (2026).
27. T. Wu, J. Jaramillo, and P. Lalanne, “Reflections on the spatial exponential growth of electromagnetic quasinormal modes,” Laser & Photonics Rev., 19, 2402133 (2025).
28. M. Abdelrahman, C. Rockstuhl, and I. Fernandez-Corbaton, “Broadband suppression of backscattering at optical frequencies using low permittivity dielectric spheres,” Sci. reports, 7, 14762 (2015).
29. S. Yi, et al. Subwavelength angle-sensing photodetectors inspired by directional hearing in small animals. Nat. Nanotechnol, 13, 1143-1147(2018).
30. C. Gigli et al., “Fundamental Limitations of Huygens’ Metasurfaces for Optical Beam Shaping,” Laser Photonics Rev. 15, 2000448 (2021).
31. Z. Jia and G. Stewart, “An analysis of the Rayleigh-Ritz method for approximating eigenspaces,” Math. Comput., 70, 637-647(2001).
32. W. Yan, R. Faggiani, and P. Lalanne, “Rigorous modal analysis of plasmonic resonances,” Phys. Rev. B 97, 205422 (2018).
33. C. Sauvan et al., “Theory of the spontaneous optical emission of nanosize photonic and plasmon resonators,” Phys. Rev. Lett. 110, 237401 (2013).
34. B. Vial et al., “Quasimodal expansion of electromagnetic fields in open two-dimensional structures,” Phys. Rev. A, 89, 023829 (2014).
35. K. Smith, A. Nixon, and D. Masiello, “Strongly coupled photonic molecules as doubly coupled oscillators,” Phys. Rev. A, 112, 063535 (2025).
36. J. Yang, H. Giessen, and P. Lalanne, “Simple analytical expression for the peak-frequency shifts of plasmonic resonances for sensing,” Nano letters, 15, 3439-3444 (2015).
37. A. Gras, P. Lalanne, and M. Duruflé, “Nonuniqueness of the quasinormal mode expansion of electromagnetic Lorentz dispersive materials,” J. Opt. Soc. Am. A 37, 1219–1228 (2020).
38. T. Wu et al., “Modal analysis of electromagnetic resonators: userguide for the MAN program,” Comput. Phys. Commun. 284,108627 (2023).
39. Q. Tao, Y. Su, Y. Zhong, H. Liu, “Efficient method for modeling large-scale arrays of optical nanoresonators based on the coupling theory of quasinormal mode,” Opt. Express, 32, 7171 (2024).
40. H. Zhao and H. Liu, “Efficient electromagnetic modeling method for large-scale arrays of optical nanoresonators on a metallic substrate based on the coupling theory of quasinormal mode,” Opt. Express 34, 16802 (2026).
41. T. Christopoulos, T. Wu, and P. Lalanne, “Modal analysis of electromagnetic resonators: MAN software expansion to 2D materials, Fano parameters, and coupled systems,” Comput. Phys. Commun. 329,110358 (2026).